\documentclass[12pt,a4paper]{article}
\usepackage{jheppub}  
\usepackage{amsfonts}    
\usepackage{dsfont}
\usepackage{pdfpages}
\usepackage{verbatim}
\usepackage{tikz}

\usepackage[utf8]{inputenc}
\usepackage[OT1]{fontenc}
\usepackage[english]{babel}

\usepackage[sc]{mathpazo}
\usepackage{amsmath,amssymb,amsfonts,mathrsfs}
\usepackage[amsmath,thmmarks]{ntheorem}

\usepackage{graphicx}
\usepackage{soul}
\usepackage{pdfpages}
\usepackage{natbib}
\usepackage{physics}
\usepackage{tensor}

\usepackage{varioref}
\usepackage{mathtools}
\usepackage[h]{esvect}
\usepackage{array}
\usepackage{listings}
\usepackage[activate]{pdfcprot}
\usepackage{booktabs}
\usepackage[mathscr]{euscript}
\usepackage{amsbsy}
\usepackage{bbm}

\usepackage[justification=centering]{caption}

\usepackage{lmodern}

\usepackage{xcolor}

\usepackage{siunitx}

\def \s= {\!\!\!=\!\!\!}

\def \ds {\partial}

\def \ag {\alpha}
\def \bg {\beta}
\def \cg {\gamma}

\def \si {\sigma}

\def \ep {\epsilon}

\def \la {\lambda}

\def \Zs {\mathbb{Z}}
\def \Rs {\mathbb{R}}
\def \Ts {\mathbb{T}}

\def \nn {\mathcal{N}}

\def \vt {\vartheta}
\def \ps {\mathfrak{psu}(1,1\vert 2)}
\def \sl {\mathfrak{sl}}
\def \slf {\mathfrak{sl}(2, \Rs)}

\def \su {\mathfrak{su}}

\newcommand{\ba}{\begin{align}}

\newcommand{\be}{\begin{equation}}
\newcommand{\ee}{\end{equation}}
\def\bd{\begin{tikzpicture}}
\def\ed{\end{tikzpicture}}
\newcommand\BigBox{\vcenter{\hbox{\scalebox{2}{$\Box$}}}}
\newcommand\bigsquare{\mathop{\BigBox}\limits}
\newcommand\partfunc[2]{\vcenter{\hbox{$\textstyle #1$}}{\bigsquare_{\textstyle #2}}}

\allowdisplaybreaks[1]

\title{Stringy CFT duals with ${\cal N}=(2,2)$ supersymmetry}

\author{Matthias R.\ Gaberdiel and Jeremy A.\ Mann} 
\affiliation{Institut f\"ur Theoretische Physik, ETH Zurich, \\
CH-8093 Z\"urich, Switzerland}
\emailAdd{gaberdiel@itp.phys.ethz.ch, mannj@student.ethz.ch}

\abstract{It was recently shown that string theory on ${\rm AdS}_3\times {\rm S}^3 \times \mathbb{T}^4$ with minimal NS-NS flux ($k=1$) is exactly dual to the symmetric orbifold of $\mathbb{T}^4$. Here we show that a similar statement also holds for the $D_n$ orbifolds of these backgrounds that have ${\cal N}=(2,2)$ supersymmetry. In this case the CFT dual is the symmetric orbifold of $\mathbb{T}^4/D_n$.}

\begin{document}

\maketitle

\makeatletter
\g@addto@macro\bfseries{\boldmath}
\makeatother

\section{Introduction} \label{sec:intro}

In \cite{Datta:2017ert}, backgrounds of the form 
\be\label{background}
{\rm AdS}_3 \times  \Bigl( {\rm S}^3 \times \mathbb{T}^4 \Bigr) / D_n \ , 
\ee
were shown to have ${\cal N}=(2,2)$ spacetime supersymmetry after orbifolding by dihedral groups $D_n$, $n=1,2,3,4,6$. Here the generators of $D_n$ act geometrically on the two $\mathbb{T}^2$'s in $\mathbb{T}^4 \cong \mathbb{T}^2 \times \mathbb{T}^2$, and the reflection generators of $D_n$ rotate the ${\rm S}^3$  by $180$ degrees. It was furthermore proposed that the CFT dual of this string background should lie on the same moduli space as 
\be\label{symorb}
{\rm Sym}_N\bigl(\mathbb{T}^4 / D_n \bigr) \ , 
\ee
where the $D_n$ action on $\mathbb{T}^4$ is the same as above. 
  
Without the $D_n$ orbifolds, the $\nn = (4,4)$ duality was originally proposed in \cite{Maldacena:1997re}, see \cite{David:2002wn} for a review, and it was recently understood from a microscopic viewpoint in \cite{Gaberdiel:2018rqv,Eberhardt:2018ouy}. To this end the string background with pure NS-NS flux was considered, in which case an exact worldsheet description is available \cite{Maldacena:2000hw,Maldacena:2000kv,Maldacena:2001km} (for the generalisation to the supersymmetric setup see also \cite{Giveon:1998ns,Israel:2003ry,Raju:2007uj,Ferreira:2017pgt}) in terms of a WZW model based on $\mathfrak{sl}(2,\mathds{R})$. It was proposed in \cite{Gaberdiel:2018rqv} that the theory with $k=1$ should be exactly dual to the symmetric orbifold of $\mathbb{T}^4$. The construction of this model in the RNS language is a bit problematic, but can be made sense of in the hybrid formalism of \cite{Berkovits:1999im}, where the worldsheet fields organise themselves (for pure NSNS flux) into a WZW model based on the superalgebra $\mathfrak{psu}(1,1|2)_k$. The latter was used in \cite{Eberhardt:2018ouy} to demonstrate an exact agreement between the spacetime spectrum of the hybrid theory and that of the symmetric orbifold of $\mathbb{T}^4$. Subsequently, it was shown in \cite{Eberhardt:2019qcl} that the operator algebra of the symmetric orbifold can also be reconstructed from the worldsheet. Again in this paper, it was noted that one may generalise the analysis to $k>1$ for which the long string spectrum of the string theory is matched with the symmetric orbifold of (${\cal N}=4$ Liouville theory) $\times \, \mathbb{T}^4$. 
\smallskip

The aim of this paper is to perform a similar analysis for the orbifolds of the form (\ref{background}). We shall mainly concentrate on the case with $k=1$, for which we expect again a direct match to the symmetric orbifold in (\ref{symorb}). The action of the $D_n$ generators on the RNS worldsheet fields was already worked out in \cite{Datta:2017ert}. The relation between these degrees of freedom and those appearing in the hybrid formalism of \cite{Berkovits:1999im} was spelled out in \cite{Eberhardt:2019qcl}, and this allows us to determine the $D_n$ action on the fields of the hybrid string.  It is then straightforward to perform an analysis similar to \cite{Eberhardt:2018ouy}, 
resulting again in an exact match of the spectra. This gives strong evidence for the duality between (\ref{background}) and (\ref{symorb}). 
\medskip

The paper is organised as follows. In Section~\ref{sec:Dnaction} we show that the $D_n$ action on the fields in the RNS formalism can be expressed in terms of rotation generators of various ${\rm SU}(2)$ symmetry groups of the background. This then allows us to translate the $D_n$ action to the hybrid formulation, see Section~\ref{sec:hybridaction}. In order to keep track of the group action on the full spectrum, we generalise the analysis of \cite{Eberhardt:2018ouy} by introducing the corresponding chemical potentials in Section~\ref{sec:chemical}. This is relatively straightforward, except for the behaviour of the ghost fields which requires some explanation (see Section~\ref{sec:3.1}). Section~\ref{sec:orbifold} is concerned with calculating the spacetime spectrum of the world-sheet orbifold for $k=1$, and we show that this reproduces indeed the symmetric orbifold spectrum of (\ref{symorb}). We explain in Section~\ref{sec:k>1} how our analysis generalises for $k>1$, for which the dual symmetric orbifold is given by (\ref{dualorb}), and we end in Section~\ref{sec:concl} with some conclusions. There is one appendix in which some aspects of the representation theory of $D_n$ are summarised. 

\section{The $D_n$ action}\label{sec:Dnaction}

Let us begin by reviewing the description of the orbifold theory in the RNS formalism. Before orbifolding the degrees of freedom of the world-sheet theory consist of 
\begin{equation}
\slf^{(1)}_k [J, \psi] \ \oplus \ \su (2)^{(1)}_k[K, \chi] \ \oplus  \ (\Ts^4)^{(1)}[\ds X, \la] \ \oplus \ \mathrm{Fock}[b,c,{\bg},{\cg}]\ ,
    \label{RNSworldsheet}
\end{equation}
where the $(1)$ superscript indicates that these are ${\cal N}=1$ superaffine models (with the notation for the relevant fields in square brackets), and the $(b,c)$ and $({\bg},{\cg})$ denote the conformal and superconformal ghosts, respectively.

The dihedral group generators act on the various world-sheet fields as follows. The bosons and fermions of the $\Ts^4$ torus transform in the fundamental representation of ${\rm SO}(4)$, and we can define the $D_n$ action on them by using that $D_n$ is naturally a subgroup of the orthogonal group in two dimensions ${\rm O}(2)$, together with 
\be
D_n \subset {\rm O}(2)_{\rm diag} \subset S\bigl( {\rm O}(2) \times {\rm O}(2) \bigr) \subset {\rm SO}(4) \ , 
\ee
where $S\bigl( {\rm O}(2) \times {\rm O}(2) \bigr)$ is the subgroup of ${\rm O}(2) \times {\rm O}(2)$ for which the product of determinants is $+1$. In terms of the representation theory of $D_n$ (that is reviewed in Appendix~\ref{app:dihedral}), this means that both the bosons and the fermions transform in 
\be\label{Dntorus}
[\ds X, \la ] \in \rho_1 \oplus \rho_1 \ . 
\ee
The action of $D_n$ on the Narain lattice $\Gamma^{4,4}$ of momentum and winding states was discussed in detail in \cite{Datta:2017ert}, see in particular Section~2.3 of that paper, and it turns out there are two inequivalent choices $D_n^{(1,2)}$ for $n=1,2,3$. Finally, the fields associated to ${\rm AdS}_3$ are invariant under $D_n$, while the generators of the sphere get rotated by $180$ degrees if the generator of $D_n$ is a reflection generator (and are invariant otherwise). 
\smallskip

For the following it will be convenient to write the above group actions in terms of `current' generators (so that we can determine the trace with the insertion of a group element in terms of the corresponding chemical potentials). Actually, it will be convenient (and sufficient) to introduce chemical potentials only for all fermionic degrees of freedom, as well as the bosonic degrees of freedom associated to ${\rm AdS}_3\times {\rm S}^3$, since the bosonic torus modes are unaffected by the transformation from the RNS formalism to the hybrid formalism (and hence can be treated as in the original RNS case). For the (fermionic) torus degrees of freedom, the relevant current algebra is $\mathfrak{so}(4)_1$ that is generated by the bilinears of the four real fermions. The corresponding zero mode algebra (that is relevant for this discussion) decomposes as 
\be\label{so4}
\mathfrak{so}(4) \cong \su(2)_{+} \oplus \su(2)_{-} \ ,
\ee
and we denote the two sets of $\su(2)$ generators by $M^a_\pm$. 
The $D_n$ action given by $\rho_1 \oplus \rho_1$, see eq.~(\ref{Dntorus}), actually takes values in ${\rm SO}(4)$, and hence can be written in terms of exponentials of $\su(2)_{+} \oplus \su(2)_-$ generators,
\begin{equation}
\rho_{\,\Ts^4} (P) = \mathrm{Ad}(e^{\pi i ( M_{+}^{1} + M_{-}^{1})})\ , \qquad \rho_{\,\Ts^4}(U) = \mathrm{Ad}(e^{\frac{4 \pi i}{n} M_{+}^3}) \ ,
\label{DnTorusFermionsInnerAutomorphism}
\end{equation}
where $U$ is the rotation generator, and $P$ the reflection generator of $D_n$. 
Here we work with the convention that 
\be
e^{\pi i M^1} = - i\,  \left( \begin{matrix} 0 & 1 \cr 1 & 0 \end{matrix} \right) \ , \qquad 
e^{\frac{4\pi i}{n} M^3} = \left(\begin{matrix} e^{\frac{2\pi i}{n}} & 0 \cr 0 & e^{-\frac{2\pi i}{n}}  \end{matrix} \right) \  ,
\ee
and it is easy to see that this describes (an equivalent representation to) $\rho_1 \oplus \rho_1$, see eq.~(\ref{eqn:DefRepDn}) and  \cite[Appendix~A]{Datta:2017ert}. Note that this construction just reflects that the fundamental representation ${\bf 4}$ of $\mathfrak{so}(4)$ corresponds to ${\bf 4} \cong ({\bf 2},{\bf 2})$ in terms of $\su(2)_+ \oplus \su(2)_-$. 

Finally, the rotation action on ${\rm S}^3$ (by $180$ degrees for the case of $P$, and trivial in the case of $U$) can be written in terms of the current generators associated to the $\su(2)^{(1)}_k$ algebra (whose global algebra we shall refer to as $\su(2)_{\rm R}$ in the following with generators $K^a$)
\begin{equation}
\rho_{\su (2)_k^{(1)}} (P) = \mathrm{Ad}(e^{\pi i K^3})\ , \qquad \rho_{\su (2)_k^{(1)}} (U) = 1\ .
\label{DnInnerAutoS3}
\end{equation}
Thus the $t^3$-valued currents (and fermions) are invariant, while the $t^\pm$-valued currents (and fermions) transform in the $\rho_-$ representation of $D_n$. Altogether the $D_n$ action on the RNS fields (except for the torus bosons) is therefore given by 
\be\label{DnActionRNS}
\rho_{\, {\rm RNS}} (P) = \mathrm{Ad}(e^{\pi i ( K^3 + M_{+}^1 + M_{-}^1)})\ , \qquad \rho_{\, {\rm RNS}}(U) = \mathrm{Ad}(e^{\frac{4 \pi i}{n} M_{+}^3}) \  . 
\ee

\subsection{Translation to the hybrid fields}\label{sec:hybridaction}

Next we want to translate these group actions to the hybrid fields. In the hybrid formalism of \cite{Berkovits:1999im}, the RNS world-sheet CFT of \eqref{RNSworldsheet} is reorganised as 
\begin{equation}
\ps_k [J,K,S] \oplus \Ts^4_{\mathrm{twisted}} [\ds X, \Psi] \oplus \mathrm{Fock}[b,c,\rho] \ ,
    \label{HybridWorldsheet}
\end{equation}
where the fermions of $\Ts^4_{\mathrm{twisted}}$ have conformal dimension $h=1$ or $h=0$ (`topologically twisted'), and $\rho$ is a boson of negative metric and background charge $Q=3$. (We are using the conventions of \cite{Eberhardt:2019qcl}). More specifically, the reformulation only affects the fermions (and the ghosts), but does not touch the (decoupled) bosonic generators of the RNS formalism. The fermions of the hybrid description can be re-expressed in terms of the RNS fields as, see Section~3.2 of \cite{Eberhardt:2019qcl}
\begin{align}
    p^{A\ag\bg} &= e^{\frac{A}{2}H_1 + \frac{\ag}{2}H_2 + \frac{\bg}{2}(H_4 + H_5) + \frac{\bg}{2} ( A \ag H_3 - \phi)}\ ,  \label{pdef}\\
    \Psi^{\mu \bg} &= e^{\frac{\mu}{2}(H_4 -H_5) +\frac{\bg}{2}(H_4 + H_5) + \bg (-\phi + \chi)}\ , \label{Psidef}
\end{align}
where $\frac{1}{2}A, \frac{1}{2}\ag, \frac{1}{2}\mu, \frac{1}{2}\bg \in \{\pm\frac{1}{2}\}$ are the spins of these fermionic fields with respect to the global $\slf \oplus \su (2)_{\rm R} \oplus \su (2)_{+} \oplus \su (2)_{-}$, respectively.
As a consequence, the $p^{A\ag\bg}$ transform in the $(\mathbf{2},\mathbf{2},\mathbf{2})$ of $\slf \oplus \su (2)_{\rm R} \oplus \su (2)_{-}$, while the $\Psi^{\mu\bg}$ transform in the $(\mathbf{2}_+,\mathbf{2}_-)$ of $\su (2)_{+} \oplus \su (2)_{-}$.\footnote{We will sometimes use the notation ${\bf 2}_\pm$ and ${\bf 2}_{\rm R}$ to indicate with respect to which $\su(2)$ algebra the relevant states transform.}

The eight fields $p^{A\ag\bg}$ can be separated into four fields $p^{A\ag} := p^{A\ag +}$ with conformal weight $h=1$, and their four conjugate fields $\theta^{A\ag} := p^{A\ag -}$ with conformal weight $h=0$. This defines four fermionic first order systems with $\la = 1$, which can be combined with the bosonic currents of $\slf_{k+2} \oplus \su (2)_{k-2}$ to produce a free field (Wakimoto) representation of $\ps_k$. In particular, the supercurrents $S^{A\ag\bg}$ transforming in the $(\mathbf{2},\mathbf{2},\mathbf{2})$ of $\slf \oplus \su (2)_{\rm R} \oplus \su (2)_{-}$ are given by
\begin{align}
S^{A\ag+} &= p^{A\ag}, \\
S^{A\ag-} &= k\ds \theta^{A\ag} +J^a c_a \tensor{(\si_a)}{^A_B} \,\theta^{B\ag} - K^a \tensor{(\si_a)}{^\ag_\bg} \,\theta^{A\bg}. \label{s3:2:3:waki_S_A_ag_minus}
\end{align}
These fields are uncharged with respect to $\mathfrak{su}(2)_+$, and hence we find from (\ref{DnTorusFermionsInnerAutomorphism}) and (\ref{DnInnerAutoS3}) that they transform under the $D_n$ action as 
\begin{equation}
    \rho_{\ps_k} (P) = \mathrm{Ad}(e^{\pi i (K^3 +M^1_{-})})\ , \qquad \rho_{\ps_k} (U) = 1\ .
    \label{DnActionPSU}
\end{equation}
\smallskip

The topologically twisted fermions $\Psi^{\mu\bg}$ transform in the $(\mathbf{2}_+,\mathbf{2}_-)$ with respect to $\su(2)_+ \oplus \su(2)_-$, 
just like their RNS counterparts, and hence their action is also described by $\rho_{\,\Ts^4}$, see eq.~\eqref{DnTorusFermionsInnerAutomorphism}. Altogether we therefore get the $D_n$ action on the hybrid fields 
\be\label{DnActionhybrid}
\rho_{\, {\rm hybrid}} (P) = \mathrm{Ad}(e^{\pi i ( K^3 + M_{+}^1 + M_{-}^1)})\ , \qquad \rho_{\, {\rm hybrid}}(U) = \mathrm{Ad}(e^{\frac{4 \pi i}{n} M_{+}^3}) \  ,
\ee
which, by construction, agrees with that on the RNS fields, see eq.~(\ref{DnActionRNS}).
Finally, while the $\rho$-ghost is expressed in terms of $\Ts^4$ degrees of freedom, 
\begin{align}\label{rhodef}
    \ds \rho = - (\ds H_4 + \ds H_5) + 2 \ds \phi - \ds \chi\ , 
\end{align}
it remains invariant under the induced $D_n$ action. (Note that $\ds H_4 + \ds H_5$ is the bilinear fermionic current associated with $M^3_-$.)

\section{Introducing chemical potentials}\label{sec:chemical}

In order to proceed it is convenient to introduce appropriate chemical potentials into the original analysis of \cite{Eberhardt:2018ouy} since this will allow us to keep track of the various group actions relatively easily. We will work with the conventions that 
\begin{equation}
\mathrm{ch}(u,v,z,t;\tau) := \mathrm{tr} \, e^{2\pi i (u M^3_{+} +  v M^3_{-} )} y^{K^3} x^{J_0^3} q^{L_0 - \frac{c}{24}}\ ,
\end{equation}
where 
\be
q = e^{2\pi i \tau} \ , \qquad y = e^{2\pi i z} \ , \qquad x= e^{2\pi i t} \ . 
\ee
In particular, the action of the $D_n$ generators $P$ and $R$ inside the trace can be absorbed into shifting the chemical potentials as\footnote{Note that since the $P$ action involves $M^1_\pm$, it is convenient to work in the basis where $M^1_\pm$ rather than $M^3_\pm$ is diagonal in $P$-twisted sectors, but for the calculation of the character this is immaterial.}
\begin{align}
P: \qquad & (u,v,z,t) \mapsto \bigl(u+\tfrac{1}{2}, v+\tfrac{1}{2}, z+\tfrac{1}{2}, t)  \label{Pchar}\\
U: \qquad & (u,v,z,t) \mapsto \bigl( u+\tfrac{2}{n}, v, z, t) \ , \label{Rchar}
\end{align}
as follows directly from eq.~(\ref{DnActionhybrid}).

We shall mainly be interested in the case where $k=1$, for which the representation theory of $\ps_k$ is very restrictive, see \cite[Section~4.2]{Eberhardt:2018ouy} for a detailed analysis. In particular, only the $j=\frac{1}{2}$ continuous representations (at the `bottom' of the continuum) are allowed, together with their spectrally flowed versions. These representations are thus labelled by the spectral flow parameter $w$, as well as $\lambda\in [0,1)$, defining the eigenvalues of $J^3_0$ modulo integers.

For the calculation of the characters (that possess many null-vectors) it will be convenient to use the free field realisation of $\ps_1$ in terms of two complex fermions and two pairs of symplectic bosons, see \cite[Section~4.5]{Eberhardt:2018ouy}. Here the symplectic boson bilinears generate $\slf_1$, the fermionic bilinears generate $\su (2)_1$, while the boson-fermion bilinears define the supercharges. The $D_n$ action given in (\ref{DnActionPSU}) can be lifted to the free fields by taking the symplectic bosons to be invariant, while the complex fermions transform as $({\bf 2}_{\rm R},{\bf 2}_-)$ with respect to  $\su(2)_{\rm R} \oplus \su (2)_-$. Then the relevant characters take the form, see \cite[Section 5]{Eberhardt:2019niq}
\begin{equation}
\mathrm{ch}_{w,\la} (v,z,t;\tau) = \sum_{m \in{\mathbb{Z}} + \lambda} q^{-mw + w^2/2}x^m \, \frac{\vt_1 (\frac{t+z + v}{2};\tau) \, \vt_1 (\frac{t-z + v}{2};\tau)}{ \eta (\tau)^4}\ .
\label{s3:4:1:TensionlessPsuCharacter}
\end{equation}

\subsection{The ghost contribution}\label{sec:3.1}

While the discussion so far is relatively straightforward, there is one subtle point we need to explain in more detail. Naively, one may have guessed that the ghosts are invariant under the $D_n$ action, but this is, in some sense, not quite correct. The basic reason for this can be read off from the DDF analysis of \cite{Eberhardt:2019qcl}.  To be specific, let us for example consider the DDF generators that correspond to the fermionic torus directions in the $-\frac{1}{2}$ picture (cf.\ eqs.~(5.8) and (5.9) of \cite{Eberhardt:2019qcl})
\be
\Lambda_r^{(-\frac{1}{2}) \mu \alpha} = k^{-\frac{1}{4}} \, \oint dz \, \Bigl( p^{+\alpha -} \, \Psi^{\mu-} e^{-\rho} \gamma^{r+\frac{1}{2}} + 
p^{-\alpha -} \, \Psi^{\mu-} e^{-\rho} \gamma^{r-\frac{1}{2}} \Bigr) \ , 
\ee
where $p^{A\alpha\beta}$, $\Psi^{\mu\beta}$ and $\rho$ were defined in eqs.~(\ref{pdef}), (\ref{Psidef}) and (\ref{rhodef}), respectively.
The original torus excitations, i.e.\ the $\Psi^{\mu\beta}$, obviously only carry charge with respect to $\su(2)_+\oplus\su(2)_-$. However, after combining with the other hybrid fields to form DDF operators (that map physical states to physical states), they acquire charge with respect to the $\mathfrak{su}(2)_{\rm R}$ symmetry coming from the ${\rm S}^3$, i.e. they now have an $\alpha$ index instead of a $\beta$ index.\footnote{The $\su(2)_{\rm R}$ becomes the R-symmetry of the spacetime CFT, and the torus fermions of the spacetime theory are indeed charged under it.} If we want to describe this effect in terms of ghosts (that eliminate the unphysical degrees of freedom), then we must take the ghosts to have some effective charge with respect to the various $\su(2)$'s (despite the fact that, on the face of it, they are uncharged with respect to any of these $\su(2)$'s.)

In order to do this more quantitatively,  we shall proceed as follows. We know that, at least for $k\geq 2$, the RNS formalism and the hybrid formalism are equivalent once physical state conditions are imposed. We also know the $\su(2)$ transformation properties (and hence those with respect to $D_n$) of the RNS and hybrid fields before imposing the physical state condition. This will allow us to deduce how the hybrid ghosts behave `effectively' with respect to these $\su(2)$ charges, and hence also under the $D_n$ action. 

To start with, recall that in the RNS formalism the world-sheet fields consist of 
\be
\slf_k^{(1)} \oplus \su(2)_k^{(1)} \oplus (\Ts^4)^{(1)} \oplus \mathrm{Fock}[b,c,\bg,\cg]\ ,
\ee
where the superscript $(1)$ indicates that these are all ${\cal N}=1$ superconformal algebras. After decoupling the fermions, imposing the physical state condition on the fermions (so as to reduce their number from $10$ to $8$), and interpreting them from the spacetime perspective --- this can be either done by using the abstruse identity, see e.g.\  the discussion in \cite[Section~2.3]{Gaberdiel:2018rqv}, or by using the DDF construction from above --- these degrees of freedom transform as 
\be\label{NSgen}
\slf_{k+2} \oplus \su(2)_{k-2} \oplus \Ts^4_{\rm bos} \oplus \mathrm{Fock}[b,c] \oplus \mathrm{Fock}[\bigl(({\bf 2}_{\rm R},{\bf 2}_+) \oplus ({\bf 2}_{\rm R},{\bf 2}_-)\bigr)\, \mathrm{fermions}] \ .
\ee
This is to be compared with the analysis in the hybrid formalism where the degrees of freedom transform as 
\be
\ps_k \oplus \Ts^4_{\rm bos} \oplus \mathrm{Fock}[b,c,\rho] \oplus \mathrm{Fock}[({\bf 2}_{+},{\bf 2}_-)\, \mathrm{fermions}] \ , 
\ee
where the last term comes from the topologically twisted fermions $\Psi^{\mu\beta}$. Using the Wakimoto representation of $\ps_k$  that was described above, see the discussion around eq.~(\ref{s3:2:3:waki_S_A_ag_minus}), the $\ps_k$ factor corresponds to 
\be
\ps_k \  \cong \ \slf_{k+2} \oplus \su(2)_{k-2}  \oplus  \mathrm{Fock}[2 \cdot  ({\bf 2}_{\rm R},{\bf 2}_{-}) \, \mathrm{fermions}] \ .
\ee  
Thus the effective ghost contribution in the hybrid formalism must remove one set of $({\bf 2}_{\rm R},{\bf 2}_{-})$ fermions and one set of $({\bf 2}_{+},{\bf 2}_{-})$ fermions, and replace it by one set of $({\bf 2}_{\rm R},{\bf 2}_{+})$ fermions, 
\be
 \{\mathrm{eff.\ ghost} \} \sim \frac{\mathrm{Fock}[\{(\mathbf{2}_{\rm R},\mathbf{2}_+)\, \mathrm{fermions}\}]}{ \mathrm{Fock}[\{(\mathbf{2}_{\rm R},\mathbf{2}_-)\, \mathrm{fermions}\}] \cdot \mathrm{Fock}[\{(\mathbf{2}_{+},\mathbf{2}_-)\, \mathrm{fermions}\}]} \ . 
 \ee
In terms of characters it must therefore take the form 
\be\label{ghostcharacter}
Z^{\mathrm{gh}} (u,v,z,t;\tau) =  \underbrace{\Bigl| \frac{\eta(\tau)^2}{\vt_1(\frac{t+z+v}{2};\tau) \vt_1(\frac{t-z+v}{2};\tau)} \Bigr|^2}_{Z^{\mathrm{gh}}_1} \cdot  \underbrace{\Bigl| \frac{\vt_1(\frac{t+z+u}{2};\tau) \vt_1(\frac{t-z+u}{2};\tau)} {\vt_1(\frac{t+u+v}{2};\tau) \vt_1(\frac{t-u+v}{2};\tau)} \Bigr|^2}_{Z^{\mathrm{gh}}_2} \ . 
\ee
Obviously, this identification is only formal since the BRST cohomology will mix all the fields together, and we cannot just extract the contribution from the additional $\rho$ ghost in this manner. However, on the level of characters this identity will be correct, and it will allow us to keep track of the $D_n$ transformation properties of the fields.

In particular, we see from this analysis that for $k=1$, for which the $\ps_1$ character is given by eq.~(\ref{s3:4:1:TensionlessPsuCharacter}), the first factor $Z^{\mathrm{gh}}_1$ of the ghost contribution (\ref{ghostcharacter}) cancels all the oscillator contributions of $\ps_1$, 
\begin{equation}\label{psugh1}
 \left( Z^{\ps_1} \cdot Z_1^{\mathrm{gh}} \right) (v,z,t;\tau)  = \abs{\sum_{m \in \Zs + \la} x^m q^{-mw + \frac{w^2}{2}}}^2.
\end{equation}
Thus we are only left with the zero mode sum that will be fixed by the mass-shell condition. As a consequence, the ${\rm AdS}_3 \times {\rm S}^3$ factor becomes `topological' for $k=1$, and all the remaining degrees of freedom come from the $\mathbb{T}^4$ part of the theory,  as already argued in \cite{Eberhardt:2018ouy}. Finally, the second factor $Z^{\mathrm{gh}}_2$ of (\ref{ghostcharacter}) makes sure that the resulting fermionic degrees of freedom have the correct charge by transmuting the topologically twisted fermions transforming as $(\mathbf{2}_{+},\mathbf{2}_-)$, into the spacetime fermions transforming as $(\mathbf{2}_{\rm R},\mathbf{2}_+)$.

\section{Calculating the orbifold}\label{sec:orbifold}

As we have explained in Section~\ref{sec:hybridaction}, the action of $D_n$ on the hybrid fields can be described in terms of group rotations, see (\ref{DnActionhybrid}) as well as (\ref{Pchar}) and (\ref{Rchar}). In order to respect these group symmetries, the same action must then also be applied to the effective ghost contribution, see eq.~(\ref{ghostcharacter}). 
As a consequence, the cancellation between $Z^{\mathrm{gh}}_1$ and the the $\ps_1$ character at level $k=1$ continues to hold also for the orbifold theory, see (\ref{psugh1}). Since the resulting expression  is independent of the chemical potentials $(u,v,z)$, it is unaffected by the $D_n$ action.\footnote{Recall that (\ref{DnActionPSU}) describes the full $D_n$ action on the $\ps$ superalgebra (including the action on the bosonic degrees of freedom).} It will therefore again be fixed by the mass-shell condition, exactly as in the case without orbifold, see also below.

The other factor of the world-sheet partition function (before orbifolding) equals 
\be
\left( Z^{\mathbb{T}^4} \cdot Z_1^{\mathrm{gh}} \right) (u,v,z,t;\tau) = Z^{\mathbb{T}^4}_{\rm bos}(\tau) \, 
Z^{\mathbb{T}^4}_{\rm fer}(u,z,t;\tau) \ , 
\ee
where $Z^{\mathbb{T}^4}_{\rm bos}$ and $Z^{\mathbb{T}^4}_{\rm fer}$ are the bosonic and fermionic contribution coming from the $\mathbb{T}^4$, respectively
\be
Z^{\mathbb{T}^4}_{\rm bos}(\tau)  = \frac{Z_\Theta(\tau)}{|\eta(\tau)|^4} \ , \qquad 
Z^{\mathbb{T}^4}_{\rm fer}(\tau)(u,z,t;\tau) = \Bigl| \frac{\vt_1(\frac{t+z+u}{2};\tau) \vt_1(\frac{t-z+u}{2};\tau)}{\eta^2(\tau)} \Bigr|^2 \ . \label{Zfer}
\ee
Here $Z_\Theta(\tau)$ is the lattice theta function of the torus (that accounts for the winding and momentum excitations), and in the second term we have used that, because of the second factor of (\ref{ghostcharacter}), the fermions transform effectively in $(\mathbf{2}_{\rm R},\mathbf{2}_+)$, see the discussion at the end of the previous section.

It is now straightforward to calculate the various orbifold contributions of the off-shell partition function. In particular, in the $(h,g)$ sector (where $h$ labels the twisted sector, and $g$ the insertion of the group element) we have 
\begin{equation}
\partfunc{h}{g} (u,v,z,t;\tau) = \abs{ \sum_{m\in\Zs+\la}x^m q^{-mw+\frac{w^2}{2}}}^2 \partfunc{h}{g}^{\Ts^4,  \mathrm{bos}}(\tau) \, \, \partfunc{h}{g}^{\Ts^4,\mathrm{ferm}}(u,z,t;\tau)\ .
\label{s4:2:ghsector_zeroandT4}
\end{equation}
The orbifold contribution coming from the bosonic degrees of freedom on the torus can be calculated directly from the $D_n$ action on the torus lattice, and this works exactly as in \cite{Datta:2017ert}. On the other hand, for the fermionic degrees of freedom we can keep track of the $D_n$ action by using (\ref{Pchar}) and (\ref{Rchar}), and this leads to 
\begin{equation}
\partfunc{h_{l,\beta}}{g_{k,\alpha}}^{\Ts^4,\, \mathrm{ferm}}(u,z,t;\tau) =    Z_{\Ts^4}^{\mathrm{ferm}} \left(u+\bigl(\frac{\beta}{2}+\frac{2l}{n}\bigr)\tau + \frac{\alpha}{2}+\frac{2k}{n}\,,\  z+ \frac{\beta}{2}\tau + \frac{\alpha}{2}\,, \, t\,; \,  \tau\right)\ , \label{FermionOrbifoldFormula}
\end{equation}
where we have labelled an arbitrary group element in $D_n$ by 
\be
g_{k,\alpha} = U^k P^\alpha \ , \qquad k = 0,1,\ldots,n-1\ ,  \quad \alpha=0,1 \ , 
\ee
and similarly for $h_{l,\beta}$. For $h=e$, i.e.\ $l=\beta=0$, this follows directly from eqs.~(\ref{Pchar}) and (\ref{Rchar}), and most of the other cases can be obtained using the modular transformation properties of these twisted twining characters,
\begin{equation}
\tau \rightarrow \frac{a\tau + b}{c\tau + d}\ , \qquad \partfunc{h}{g} \mapsto \partfunc{h^a g^b}{h^cg^d} \ ,
\end{equation}
together with the modular properties of (\ref{Zfer}); in particular we need the identities
\be
\begin{array}{rclrcl}
\vt_1\bigl(\frac{z}{\tau};-\frac{1}{\tau}\bigr) & = & -i \,\sqrt{-i\tau} \, e^{ \frac{\pi i z^2}{\tau}}\, \vt_1(z;\tau)  \qquad \qquad 
& \eta\bigl(-\frac{1}{\tau}\bigr) & = &  \sqrt{-i\tau} \, \eta(\tau) \ , \\[4pt] 
\vt_1(z;\tau+1) & = &  e^{i\frac{\pi}{4}} \, \vt_1(z;\tau)  \qquad 
& \eta(\tau+1) & = & e^{i\frac{\pi}{12}} \, \eta(\tau) \ , 
\end{array}
\ee
see \cite[Appendix~F]{Eberhardt:2018ouy} for our conventions. 
In fact, there is only one ${\rm SL}(2,\mathbb{Z})$ orbit of sectors that is not fixed by this argument, containing the representatives
\be
\partfunc{P}{U^k}^{\Ts^4,\, \mathrm{ferm}}(u,z,t;\tau) = Z_{\Ts^4}^{\mathrm{ferm}} \left(u+\frac{\tau}{2}+ \frac{2k}{n}\,,\  z+ \frac{\tau}{2}\,, \, t\,; \,  \tau\right)\ .
\ee
The expression for this sector can be obtained by noting that in the $P$-twisted sector two of the four fermions are half-integer moded (while the other two have integer modes), and that the $U^k$ generators act as before. 

The final step consists of imposing the physical state conditions $L_0 = 0 = \bar{L}_0$, i.e.\ picking out the term $q^0 \bar{q}^{\, 0}$ in eq.~(\ref{s4:2:ghsector_zeroandT4}). As in \cite{Eberhardt:2018ouy}, only one term in the sum survives, namely the one for which 
\be
m = \frac{h^{\Ts^4}}{w} + \frac{w}{2} \ , 
\ee
and similarly for the right-movers. Thus the $(h,g)$ sector of the string partition function becomes 
\be
Z_{\rm string}^{(h,g)} (u,z;t) = \sum_{w=1}^{\infty} x^{\frac{w}{2}} \bar{x}^{\frac{w}{2}} \, \partfunc{h}{g}^{\Ts^4,  \mathrm{bos}}\bigl(\frac{t}{w}\bigr) \, \, \partfunc{h}{g}^{\Ts^4,\mathrm{ferm}}\bigl(u,z,t;\frac{t}{w}\bigr) ' \ ,
\ee
where the prime indicates that only those states contribute for which $m-\bar{m}\in\mathbb{Z}$. Using the same theta function identitites as in 
\cite[eq.~(5.8)]{Eberhardt:2018ouy},  we can then finally rewrite this as\footnote{Here the symbols $\mathrm{R}$ and $\mathrm{NS}$ describe the moding of the fermions before considering the $h$-twisted sector, i.e.\ the twisting by $h$ changes the moding of the fermions relative to an integer moding (for the case of $\mathrm{R}$) and a half-integer moding (for the case of $\mathrm{NS}$).}
\be
Z_{\rm string}^{(h,g)} (u,z;t) =
\sum_{w \in 2\mathbb{N}} \abs{x^{\frac{w}{4}}}^2 \partfunc{h}{g}^{\mathbb{T}^4,\mathrm{R}}\bigl(u,z; \frac{t}{w}\bigr) + \sum_{w \in 2\mathbb{N}-1} \abs{x^{\frac{w}{4}}}^2 \partfunc{h}{g}^{\mathbb{T}^4,\mathrm{NS}} \bigl(u,z; \frac{t}{w}\bigr)\ .
\label{s4:2:3:FullStringPartfunc_gh}
\ee
This agrees exactly with the single particle (single cycle) part of the partition function of the symmetric orbifold of $\Ts^4 /D_n$, where $w$ describes the length of the single cycle. 

\section{Some comments about $k>1$}\label{sec:k>1}

The analysis for $k>1$ actually works very similarly. In that case we can directly use the RNS description, and there is no need to introduce the hybrid fields. There are no null-vectors for $\sl(2,\mathbb{R})_k$, so the underlying characters are the Verma module characters. If we concentrate on the long string sector, the analysis of \cite{Eberhardt:2019qcl} (together with the refinement explained in the previous section) goes through essentially unmodified, and the spacetime spectrum turns out to match exactly with 
\be\label{dualorb}
\mathrm{Sym}_N \Bigl(  \bigl[  (\hbox{${\cal N}=4$ Liouville at $c=6(k-1)$}) \times \mathbb{T}^4  \bigr] / D_n \Bigr) \ . 
\ee
It remains to explain though how $D_n$ acts on the seed theory. To this end we recall the description of the world-sheet degrees of freedom from (\ref{NSgen}),
\be\label{NSgen1}
\slf_{k+2} \oplus \su(2)_{k-2} \oplus \Ts^4_{\rm bos} \oplus \mathrm{Fock}[b,c] \oplus \mathrm{Fock}[\bigl(({\bf 2}_{\rm R},{\bf 2}_+) \oplus ({\bf 2}_{\rm R},{\bf 2}_-)\bigr)\, \mathrm{fermions}] \ .
\ee
The fermions in the $(\mathbf{2}_{\rm R},\mathbf{2}_+)$ combine with the bosons from $\Ts^4_{\rm bos}$ to give the supersymmetric $\Ts^4$ theory, on which $D_n$ acts as in (\ref{Dntorus}).  The $(b,c)$ ghosts  cancel two (of the three) bosonic degrees of freedom from $\slf_{k+2}$, and thus the remaining degrees of freedom are 
\be\label{5.3}
\hbox{free boson} \oplus \su(2)_{k-2}  \oplus \mathrm{Fock}[({\bf 2}_{\rm R},{\bf 2}_-) \, \mathrm{fermions}] \ ,
\ee
which just give rise to $\hbox{${\cal N}=4$ Liouville at $c=6(k-1)$}$, see the discussion in \cite[Section~6.2]{Eberhardt:2019qcl}. They transform under $D_n$ as follows. First the free boson arises from $\slf_{k+2}$, and hence is invariant under $D_n$. On the remaining degrees of freedom, the action of $D_n$ can be read off from (\ref{DnActionRNS}): the rotation generator $U$ of $D_n$ acts trivially (since none of these fields are charged under $\su(2)_+$), while the reflection generator $P$ rotates $\su(2)_{k-2}$ by $180$ degrees (one of the currents, say $K^3$, is invariant, while the other two pick up a sign). As regards the fermions, we note that the reflection generator is embedded diagonally into $\su(2)_{\rm R}\oplus\su(2)_{-}$, see eq.~(\ref{DnActionRNS}). The fermions in $({\bf 2}_{\rm R},{\bf 2}_-)$ transform in the tensor product ${\bf 2} \otimes {\bf 2} = {\bf 3} \oplus {\bf 1}$ with respect to this diagonal $\su(2)$, and hence their transformation property coincides with the geometric $D_n$ action on ${\rm S}^3$. (This is to say that $3$ of the four fermions transform exactly as the currents $K^a$, while the remaining fermion is invariant.) In particular, we can combine the spacetime fields of (\ref{5.3}) into 
\be\label{su2orb}
\su(2)^{(1)}_{k}  \oplus \mathfrak{u}(1)^{(1)} \ , 
\ee
where the superscript $(1)$ refers to the ${\cal N}=1$ superconformal affine algebra (for which the fermions transform in the adjoint representation). The $P$ generator of $D_n$ then acts by rotating the $\su(2)^{(1)}_{k}$ factor (including the fermions) by $180$ degrees, while leaving $\mathfrak{u}(1)^{(1)}$ invariant.\footnote{This agrees with eq.~(\ref{DnInnerAutoS3}), where now $\su(2)^{(1)}_{k}$ is part of the spacetime Liouville theory.} This then defines the $D_n$ action on the ${\cal N}=4$ Liouville factor in (\ref{dualorb}), and hence specifies the dual CFT.

This analysis applies irrespective of whether $k$ is even or odd. This is to be compared with the original analysis of \cite{Datta:2017ert}, where the duality only worked for $k$ odd since the $\mathfrak{u}(1)$ charge quantisation did not match between world-sheet and dual CFT. The reason why things are different here is that, unlike the case considered in \cite{Datta:2017ert}, the symmetric orbifold of the spacetime theory now also contains a Liouville factor which, in particular, includes the $\mathfrak{su}(2)_{k}^{(1)}$ algebra of eq.~(\ref{su2orb}). As we have just seen, this algebra is also orbifolded, thus providing an extra contribution to the charge quantisation in the twisted sector. As a consequence the charges between the two descriptions now match irrespective of whether $k$ is even or odd.\footnote{We thank Lorenz Eberhardt for a useful discussion about this issue.}
\smallskip

Finally, for $k>1$ the DDF operators can be directly constructed in the RNS formalism \cite{Giveon:1998ns,Eberhardt:2019qcl}, and their $D_n$ transformation properties follow directly from those of the RNS fields. As in \cite{Eberhardt:2019qcl}, see in particular Sections~2.5 and 2.6, the relevant modes will be $w$-fractionally moded in the $w$-spectrally flowed sector, exactly as one should expect for the $w$-cycle twisted sector of the dual CFT. As a result, the operator algebra of the symmetric orbifold (\ref{dualorb}) can also be reproduced from the world-sheet. It also seems to imply that the spacetime theory is supersymmetric for any value of $k$, in contradiction to what was argued in \cite{Datta:2017ert}.

\section{Conclusions}\label{sec:concl}

In this paper we have shown that the duality between string theory on  ${\rm AdS}_3 \times {\rm S}^3 \times \mathbb{T}^4$ with minimal NSNS flux ($k=1$) on the one hand, and the symmetric orbifold of $\mathbb{T}^4$ \cite{Eberhardt:2018ouy} on the other hand, extends also to the ${\cal N}=(2,2)$ supersymmetric $D_n$ orbifolds of these same backgrounds studied in \cite{Datta:2017ert}. Our results give strong support to the duality proposal of \cite{Datta:2017ert}, and they also demonstrate that the techniques of \cite{Eberhardt:2018ouy} are more widely applicable. 

It would be interesting to show that a similar analysis can be done for the ${\cal N}=(3,3)$ orbifold of \cite{Eberhardt:2018sce}; this should follow from the duality for ${\rm AdS}_3 \times {\rm S}^3 \times {\rm S}^3 \times {\rm S}^1$ that was first proposed in \cite{Eberhardt:2017pty} and then derived microscopically in \cite{Eberhardt:2019niq}. 

More fundamentally, it would be helpful to make these dualities more manifest. In this context it is curious to note that the Drinfel'd-Sokolov (or quantum Hamiltonian) reduction of the $\ps_k$ supergroup WZW model leads exactly to the same $\nn = 4$ Liouville theory (including its central charge) that appears for $k>1$ in \eqref{dualorb}, see e.g.~\cite{Ito92}, in particular the discussion around equation (19) of that paper with reference to table 1.\footnote{Incidentally, an analogous statement holds for the Drinfel'd-Sokolov reduction of $\mathfrak{d}(2,1;\ag)$ (also given in \cite{Ito92}), and the Liouville factor in the ${\rm AdS}_3 \times {\rm S}^3 \times {\rm S}^3 \times {\rm S}^1$ string spectrum of \cite{Eberhardt:2019niq}.} If one could express the BRST charge of these Drinfel'd-Sokolov reductions in terms of the worldsheet BRST operator, this should lead to a more conceptual (and less background dependent) derivation of these dualities. It should also allow for a more direct derivation of the ``effective ghost" contribution of Section~\ref{sec:3.1}.

\section*{Acknowledgements}  

We thank Lorenz Eberhardt for many useful discussions. The results of this paper are largely based on the Master thesis of one of us (JAM). We gratefully acknowledge the support of the NCCR SwissMAP that is funded by the Swiss National Science Foundation.

\appendix

\section{The dihedral group and its representations}\label{app:dihedral}

The dihedral groups $D_n$, $n \in \Zs_{>0}$, can be defined as subgroups of ${\rm O}(2)$. The orthogonal group has two connected components which we can write as \pagebreak
\begin{align}
{\rm O}(2) &\equiv {\rm SO}(2) \cup P\cdot {\rm SO}(2),   \nonumber\\
&= \{U(\theta)\}_{\theta \in [0,2\pi)} \cup \{PU(\theta)\}_{\theta \in [0,2\pi)} \ , 
\label{c:1:1:O2}
\end{align}
where $U(\theta)$ denotes a rotation by $\theta$, while $P$ is the reflection along the $y$-axis, say. The generators $P$ and $U(\theta)$ satisfy the relation $PU(\theta) = U(2\pi -\theta) P$, and they act on the $2$-dimensional defining representation $\rho_{\rm f}$ (written in a complex basis) as 
\begin{align}
\rho_{\rm f}(P) =  - \begin{pmatrix} 0 & 1 \\ 1 & 0 \end{pmatrix}\ , \qquad 
\rho_{\rm f}(U(\theta)) = \begin{pmatrix} e^{i\theta}  & 0 \\ 0 & e^{-i\theta} \end{pmatrix}\ , \quad \theta \in [0,2\pi)\ . 
\label{eqn:DefRepO2}
\end{align}
The discrete subgroup obtained by restricting the rotations to $U^k$, $U=U(\frac{2\pi }{n})$, defines the dihedral group $D_n$, with a defining representation $\rho_1$ given by  
\begin{equation}
\rho_1(P) :=  - \begin{pmatrix} 0 & 1 \\ 1 & 0 \end{pmatrix}\ , \qquad \rho_1(U):= \begin{pmatrix} e^{\frac{2\pi i}{n}} & 0 \\ 0 & e^{-\frac{2\pi i}{n}} \end{pmatrix}\ .
\label{eqn:DefRepDn}
\end{equation}
The structure of the group depends a little bit on whether $n$ is even or odd. 

\subsection{$D_n$ with $n=2p+1$}

For $n$ odd the conjugacy classes are described in Table~\ref{tab:D_2p+1}, and the irreducible representations are the $1$-dimensional representations $\rho_\pm$,  together with the $2$-dimensional representations $\rho_j$, $j=1,\ldots, p$. More explicitly, they are defined by 
\be
\rho_{\ep}(P) = \ep \ , \qquad \rho_{\ep}(U) = 1 \ ,  \qquad \ep = \pm \ , 
\ee
and 
\be
\rho_{j}(P) = (-)^{j+1}\, \rho_1 (P)\ , \qquad \rho_j (U) = \rho_1(U)^j\ ,  \qquad j=1,\ldots,p \ . 
\ee

\begin{table}[htb]
\center
\begin{tabular}{l l r }
\toprule
Conjugacy Class $[g]$ & Elements of $[g]$ & Centralizer of $g$ \\
\midrule
$[1]$ & 1 & $D_{2p+1}$ \\
$[U^k]_{k=1,\dots,p}$ & $U^k,U^{2p+1-k}$ & $\mathbb{Z}_{2p+1} (U)$ \\
$[P]$ & $(U^l P)_{l=0,1,\ldots,2p}$ & $\mathbb{Z}_2 (P)$ \\
\bottomrule
\end{tabular}
\caption{Conjugate orbits and stabilizers of $D_n$ for $n$ odd}
\label{tab:D_2p+1}
\end{table}

\subsection{$D_n$ with $n=2p$}

In this case, the conjugacy classes are spelled out in Table~\ref{tab:D_2p}, and the irreducible representations consist now of four $1$-dimensional representations 
\be
\rho_{\ep \eta}(P) = \ep \ , \qquad \rho_{\ep \eta}(U) = \eta \ ,  \qquad \ep, \eta = \pm \ , 
\ee
and $(p-1)$ $2$-dimensional representations 
\be
\rho_{j}(P) = (-)^{j+1}\, \rho_1 (P)\ , \qquad \rho_j (U) = \rho_1(U)^j\ ,  \qquad j=1,\ldots,p-1 \ . 
\ee

\begin{table}[htb]
\center
\begin{tabular}{l l r }
\toprule
Conjugacy Class $[g]$ & Elements of $[g]$ & Centralizer of $g$ \\
\midrule
$[1]$ & 1 & $D_{2p}$ \\
$[U^p]$ & $U^p$ & $D_{2p}$ \\
$[U^k]_{k=1,\ldots,p-1}$ & $U^k,U^{2p-k}$ & $\mathbb{Z}_{2p} (U)$ \\
$[P]$ & $(U^{2k} P)_{k=1,\ldots,p}$ & $D_2 (U, U^p)$ \\
$[UP]$ & $(U^{2k+1} P)_{k=1,\ldots,p}$ & $D_2 (U P, U^p)$ \\
\bottomrule
\end{tabular}
\caption{Conjugate orbits and stabilizers of $D_n$ for $n$ even.}
\label{tab:D_2p}
\end{table}

\end{document}